\documentstyle[epsfig]{elsart}

\begin{document}
\begin{frontmatter}
\title{Volatility distribution in the S\&P500 Stock Index}

\author[cps]{Pierre Cizeau}
\author[cps]{Yanhui Liu} 
\author[cps]{Martin Meyer}
\author[bih,cps]{C.-K. Peng} 
\author[cps]{H.~Eugene Stanley}

\address[cps]{Center for Polymer Studies and Department of Physics,
Boston University, Boston, MA 02215, USA}

\address[bih]{Harvard Medical School, Beth Israel Deaconess Medical
Center, Boston, MA 02215, USA}

\date{August 15, 1997}

\maketitle

\begin{abstract}  
We study the volatility of the S\&P500 stock index from 1984 to 1996
and find that the volatility distribution  can be very well described
by a log-normal function. Further, using detrended fluctuation analysis
we show that the volatility is power-law correlated with Hurst exponent
$\alpha\cong0.9$. 
\end{abstract}
\end{frontmatter}

\bigskip

The volatility is a measure of the mean fluctuation of a market price
over a certain time interval $T$. The volatility is of practical
importance since it quantifies the risk related to assets
\cite{Bouchaud94}. Unlike price changes that are correlated only on very
short time scales \cite{Fama70} (a few minutes), the absolute values of
price changes (which are closely related to the volatility) show
correlations on time scales up to many years
\cite{Ding93,Dacorogna93,Liu97}. 

Here we study in detail the volatility of the S\&P 500 index of the New
York stock exchange, which represents the stocks of the 500 largest
U.S. companies. Our study is based on a data set over 13 years from
January 1984 to December 1996 reported at least every minute (these
data extend by 7 years the data set previously analyzed in
\cite{Mantegna95}). 

We calculate the logarithmic increments  

\begin{equation}
G(t)\equiv\log_e Z(t+\Delta t) -\log_e Z(t)\;, 
\end{equation}
where $Z(t)$ denotes the index at time $t$ and $\Delta t$ is the time
lag; $G(t)$ is the relative price change $\Delta Z/Z$ in the limit
$\Delta t\to0$. Here we set $\Delta t=30\,$min, well above the
correlation time of the price increments; we obtain similar results 
for other choices of $\Delta t$ (larger than the correlation time). 

Over the day, the market activity shows a strong ``U-shape'' dependence
with high activity in the morning and in the afternoon and much lower
activity over noon.  To remove artificial correlations resulting from
this intra-day pattern of the volatility
\cite{Wood85,Harris86,Admati88,Ekman92}, we analyze the normalized function

\begin{equation}
g(t)\equiv G(t)/A(t)\;,
\end{equation}
where $A(t)$ is the mean value of $|G(t)|$ at the same time of the day
averaged over all days of the data set.

We obtain the volatility at a given time by averaging $|g(t)|$
over a time window $T=n\cdot\Delta t$ with some integer $n$,

\begin{equation}
v_T(t)\equiv{1\over n}\sum_{t'=t}^{t+n-1}|g(t')|\;.
\end{equation}

\begin{figure}[hbt]
\epsfxsize=14.5cm\vspace{10mm}
\epsffile{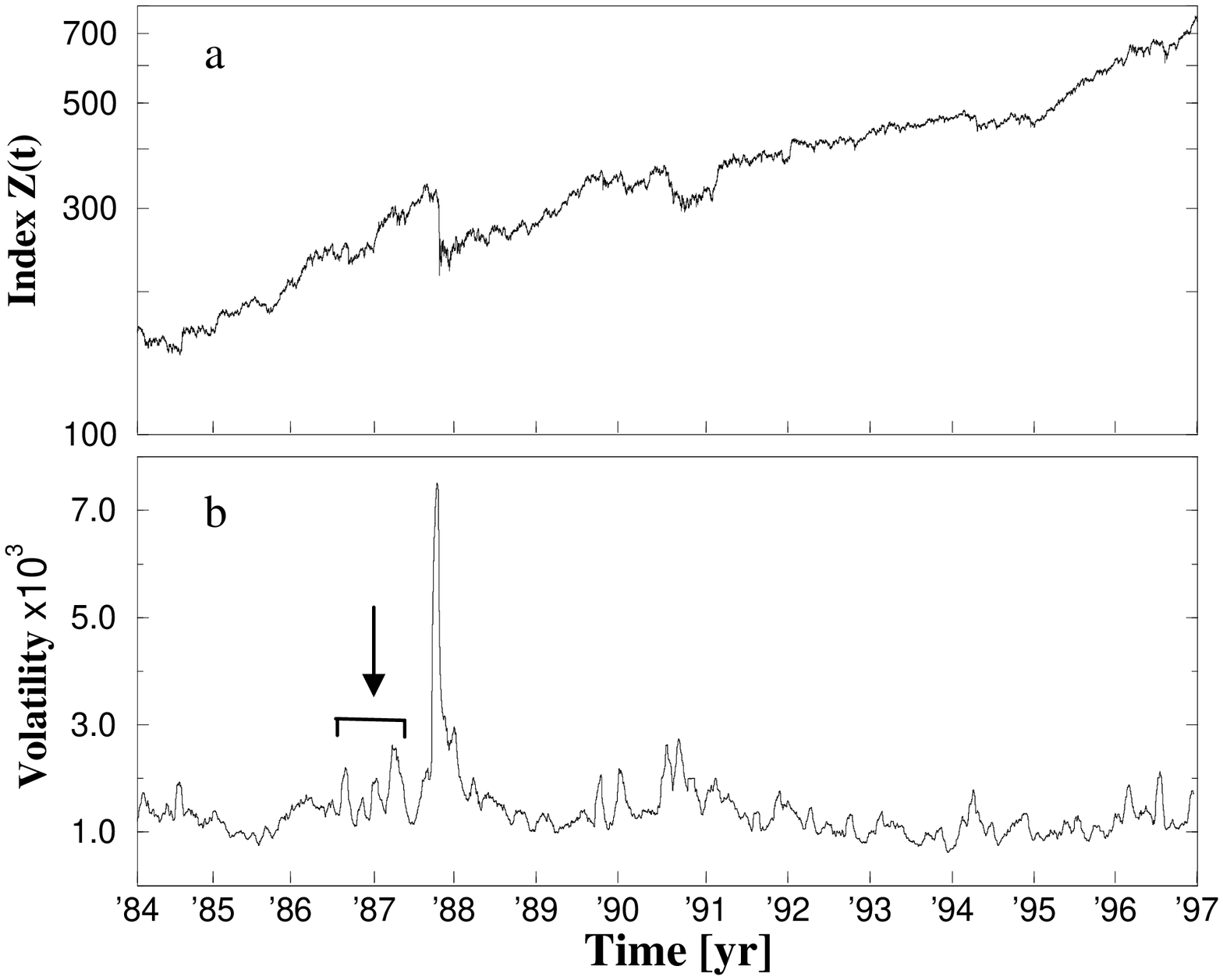}\vspace{-10mm}
\caption{ (a) Raw data analyzed: The S\&P 500 index Z(t) for the 13-year
period 1 Jan 1984 - 31 Dec 1996 at interval of 1 min. Note the large
fluctuations, such as that on 19 Oct 1987 (``black Monday''). (b)
Volatility $v_{T}(t)$ with T=1mon (8190min) and time lag 30min. The
precursors of the '87 crash are indicated by arrows.
}
\label{raw}
\end{figure}

Figure \ref{raw} shows (a) the S\&P500 index and (b) the signal
$v_T(t)$ for a long averaging window $T=8190\,$min (about 1$\,$month).
The volatility fluctuates strongly showing a marked maximum for the '87
crash. Generally periods of high volatility are not independent but
tend to ``cluster''. This clustering is especially marked around the
'87 crash, which is accompanied by precursors (possibly related to the
oscillatory patterns postulated in \cite{Sornette96}). Clustering occurs
also in other periods (e.g.~in the second  half of '90), while there
are extended periods where  the volatility remains at a rather low
level (e.g.~the years  of '85 and '93).

\begin{figure}[hbt]
\begin{center}
\begin{minipage}{12cm}
\epsfxsize=12cm
\epsffile{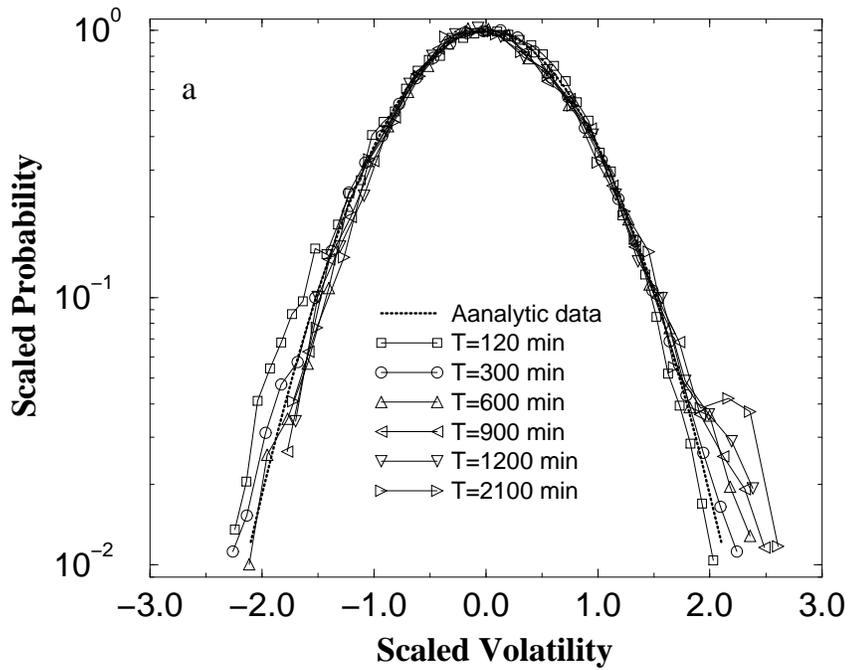}
\epsfxsize=12cm
\epsffile{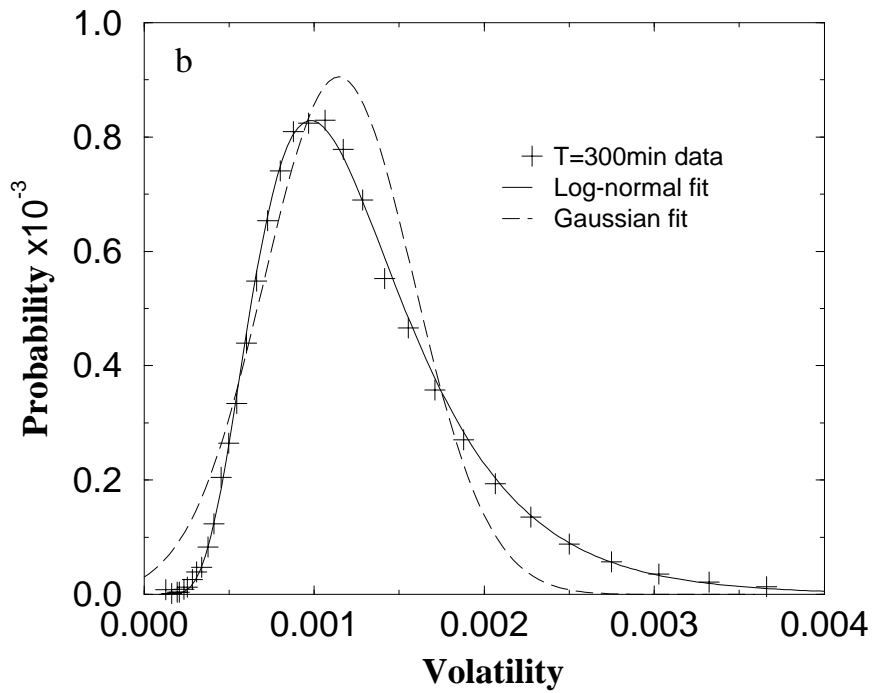}
\end{minipage}
\end{center}
\caption{ (a) The volatility distribution for different window sizes T
in scaled form, $\protect\sqrt{\mu}\exp(a+\mu/4)P(v_{T})$ as a function
of  $(\ln(v_{T})-a)/\protect\sqrt{\pi\mu}$,  where $a$ and $\mu$ are
the mean and the width  on a logarithmic scale. By the scaling, all
curves collapse to the log-normal form with $a=0$ and $\mu=-1$,
$\exp(-(\ln x)^2)$ (dotted line). (b) Comparison of the best log-normal
and Gaussian fits for the 300min data.}
\label{lognormal}
\end{figure}

Fig.~\ref{lognormal}a shows the scaled probability distribution
$P(v_T)$ for several values of $T$. The data for different averaging
windows collapse to one curve. Remarkably, the scaling form is
log-normal, not Gaussian. In the limit of very long averaging times,
one expects that $P(v_T)$ becomes Gaussian, since the central limit
theorem holds also for correlated series \cite{Beran94}, with a slower
convergence than for non-correlated processes \cite{Potters97,Cont97}.
For the times considered here, however, a log-normal fits the data
better than a Gaussian, as is evident in Fig.~\ref{lognormal}b which
compares the best log-normal fit with the best Gaussian fit for data
averaged over a $300\,$min window. 

\begin{figure}[hbt] 
\begin{center}
\begin{minipage}{12cm}
\epsfxsize=12cm
\epsffile{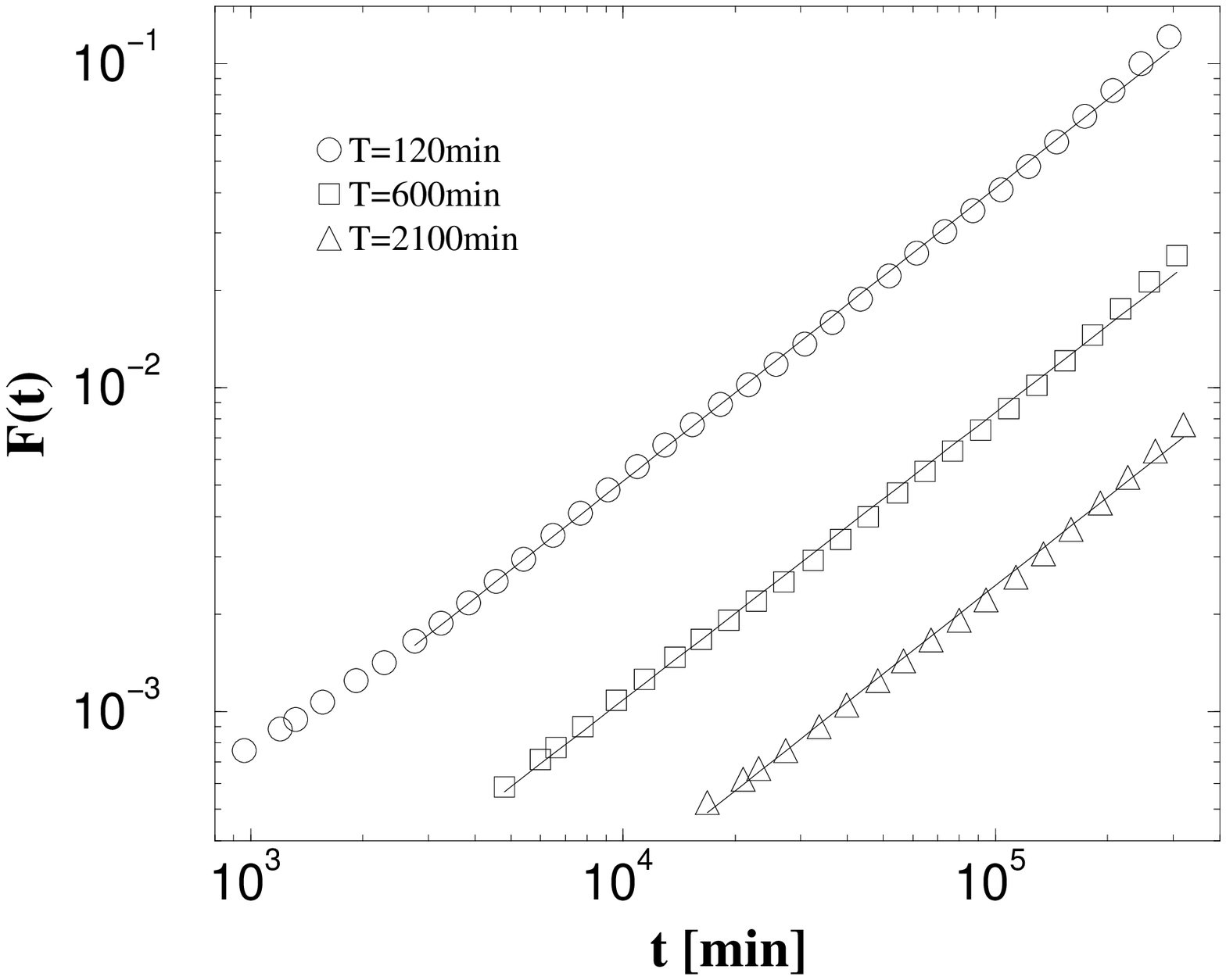}
\end{minipage}
\end{center}
\caption{ The fluctuation $F(t)$ of volatility
$v_{T}(t)$ with T=120$\,$min, 600$\,$min, and 2100$\,$min, calculated
using   detrended fluctuation analysis (DFA) \protect\cite{Peng94}.  To
implement the DFA method, we integrate $v_T(t)$ once; then we determine
the fluctuations $F(t)$ of the integrated signal around the best linear
fit in a time window of size $t$.  The lines are the best power-law
fits with exponents $\alpha=$0.91, 0.89, and 0.91.} 
\label{DFA}
\end{figure}

The correlations found in Fig.\ref{raw}b can be accurately quantified 
by detrended fluctuation analysis \cite{Peng94}. The analysis reveals
power-law behavior independent of the $T$ value chosen (Fig.~\ref{DFA})
with an exponent $\alpha\cong0.9$ in agreement with the value found
for the absolute price increments \cite{Liu97}.

To account for the time dependence of the volatility and its
correlations, ARCH \cite{Engle82}, GARCH \cite{Bollerslev86} models and
related approaches \cite{Granger96} have been developed, which assume
that the volatility depends on time and on the past evolution of the
index. It may be worthwhile to test models also for the volatility
distribution $P(v_T)$.

In summary, we have found that the probability distribution of the
S\&P500 volatility can be well described by a log-normal function and
that the volatility shows power-law correlations with Hurst exponent
$\alpha\cong0.9$. The log-normal shape of the distribution is
consistent with a multiplicative process \cite{Bunde96} for the
volatility \cite{Gasghaie96}. However, a multiplicative behavior would
be surprising, because efficient market theories \cite{Fama70} assume
that the price changes, $G(t)$, are caused by incoming new informations
about an asset. Since such information-induced price changes are
additive in $G(t)$, they should not give rise to multiplicative
behavior of the volatility.

\section*{Acknowledgments}\vspace{-5mm}
 
We thank S.~Zapperi for crucial assistance in the early stage of this
work, P.~Gopikrishnan, S.~Havlin, and R.~Mantegna for very helpful discussions, and
DFG, NIH, and NSF for financial support.

\pagebreak

\pagebreak

\begin{thebibliography}{[99]}

\bibitem{Bouchaud94} J.-P. Bouchaud and D. Sornette, {\it J.~Phys.I
France} {\bf 4} (1994) 863 .

\bibitem{Fama70} E.-F. Fama, {\it J.~Finance} {\bf 25} (1970) 383.

\bibitem{Ding93} Z. Ding, C.~W.~J. Granger  and R.~F. Engle, {\it J.
Empirical Finance} {\bf 1} (1993) 83.

\bibitem{Dacorogna93} M.~M. Dacorogna, U.~A. Muller, R.~J. Nagler,
R.~B. Olsen and O.~V. Pictet, {\it J. Int. Money and Finance}
{\bf 12} (1993) 413.

\bibitem{Liu97} Y. Liu, P. Cizeau, M. Meyer, C.-K.~Peng, and H. E.
Stanley, {\it Physica} A {\bf 245} (Nov.~1997), xxx, in press.

\bibitem{Mantegna95}  R.~N. Mantegna and H.~E. Stanley, {\it Nature}
{\bf 376} (1995) 46; {\bf 383} (1996) 587; {\it Physica} A {\bf 239}
(1997) 255.

\bibitem{Wood85} R.~A. Wood, T.~H. McInish and J.~K. Ord, {\it J. of
Finance} {\bf 40} (1985) 723.

\bibitem{Harris86} L. Harris, {\it J. of Financial Economics} {\bf 16}
(1986) 99.

\bibitem{Admati88} A. Admati and P. Pfleiderer, {\it Review of Financial
Studies} {\bf 1} (1988) 3.

\bibitem{Ekman92} P.~D. Ekman, {\it The Journal of Futures
Markets}, Vol. {\bf 12}, No. {\bf 4} (1992) 365.

\bibitem {Sornette96} D. Sornette, A. Johansen, and J.-P. Bouchaud, {\it
J.~Phys.~I France} {\bf 6} (1996) 167.

\bibitem{Beran94} J. Beran, {\it Statistics for Long-Memory Processes}
(Chapman \& Hall, NY, 1994).

\bibitem{Potters97} M.~Potters, R.~Cont, and J.-P.~Bouchaud,
cond-mat/9609172 

\bibitem{Cont97} R. Cont, cond-mat/9705075

\bibitem{Peng94} C.-K. Peng, S.~V. Buldyrev, S. Havlin, M. Simons,
H.~E. Stanley and A.~L. Goldberger, {\it Phys. Rev.} E {\bf 49} (1994)
1684. 

\bibitem{Engle82} R.F. Engle, {\it Econometrica} {\bf 50} (1982) 987.

\bibitem {Bollerslev86} T.~Bollerslev, {\it J. Econometrics} {\bf 31}
 (1986) 307.

\bibitem{Granger96} C.~W.~J. Granger and Z. Ding, {\it J. Econometrics} {\bf
73} (1996) 61.

\bibitem{Bunde96} 
A.~Bunde and S.~Havlin, in {\it Fractals and Disordered Systems},
ed. by A.~Bunde and S.~Havlin, $2^{\mbox{\tiny nd}}$ ed., 
(Springer, Heidelberg 1996).

\bibitem{Gasghaie96} S.~Ghashgaie, W.~Breymann, J.~Peinke, P. Talkner,
and Y.~Dodge, {\it Nature} {\bf 381} (1996) 767. 

\end{thebibliography}
\end{document}